\documentclass[pdftex,twocolumn,epjc3]{svjour3}          

\pdfoutput=1

\RequirePackage[T1]{fontenc}

\smartqed  

\RequirePackage{graphicx}
\RequirePackage{mathptmx}      
\RequirePackage{flushend}
\RequirePackage[numbers,sort&compress]{natbib}
\RequirePackage[colorlinks,citecolor=blue,urlcolor=blue,linkcolor=blue]{hyperref}
\RequirePackage{booktabs}
\journalname{Eur. Phys. J. C}

\usepackage{rotating}
\usepackage{amssymb,amsmath,array}
\usepackage{units}
\usepackage{heppennames2}
\usepackage{subfig}
\DeclareGraphicsExtensions{.pdf, .png}

\begin{document}

\newcommand{\epem}{\ensuremath{\mathrm{\Pep\Pem}}\xspace}
\newcommand{\abinv}{\ensuremath{\mathrm{ab}^{-1}}\xspace}
\newcommand{\pT}{\ensuremath{p_\mathrm{T}}\xspace}
\newcommand{\sqrts}{\ensuremath{\sqrt{s}}\xspace}

\newcommand{\gghadrons}{\ensuremath{\PGg\PGg \rightarrow \mathrm{hadrons}}\xspace}
\newcommand{\clicsid}{CLIC\_SiD\xspace}
\newcommand{\micron}{\ensuremath{\upmu\mathrm{m}}}
\newcommand{\radlen}{\ensuremath{X_{0}}\xspace}
\newcommand{\radlenfrac}{\ensuremath{X/X_{0}}\xspace}
\newcommand{\nuclen}{\ensuremath{\lambda_{\mathrm{I}}}\xspace}
\newcommand{\degrees}{\ensuremath{^{\circ}}\xspace}
\newcommand{\rmlad}{\ensuremath{_{\mathrm{ladder}}}}
\newcommand{\mm}[1]{\ensuremath{_{\mathrm{#1}}~\mathrm{[mm]}}}
\newcommand{\mic}[1]{\ensuremath{_{\mathrm{#1}}~\mathrm{[\micron]}}}
\newcommand{\mumu}{\ensuremath{\PGm\PGm}\xspace}
\newcommand{\nuenuebar}{\ensuremath{\PGne\PAGne}\xspace} 
\newcommand{\mpmm}{\ensuremath{\PGmp\PGmm}\xspace}  
\newcommand{\nunubar}{\ensuremath{\PGn\PAGn}\xspace}   
\newcommand{\tptm}{\ensuremath{\PGtp\PGtm}\xspace} 
\newcommand{\gamgam}{\ensuremath{\upgamma\upgamma}\xspace}
\newcommand{\ww}{\ensuremath{\PWp\PWm}\xspace} 
\newcommand{\zz}{\ensuremath{\PZ\PZ}\xspace} 
\newcommand{\wwz}{\ensuremath{\PWp\PWm\PZ}\xspace} 
\newcommand{\zzz}{\ensuremath{\PZ\PZ\PZ}\xspace} 
\newcommand{\qq}{\ensuremath{\PQq\PAQq}\xspace} 
\newcommand{\zhsm}{\ensuremath{\PH\PZ}\xspace}  
\newcommand{\hbb}{\mbox{\ensuremath{\PH\to \PQb \PAQb}}\xspace}
\newcommand{\hcc}{\mbox{\ensuremath{\PH\to \PQc \PAQc}}\xspace}
\newcommand{\hmumu}{\mbox{\ensuremath{\PH\to \mpmm}}\xspace}
\newcommand{\guineapig}{\textsc{GuineaPig}\xspace}
\newcommand{\mokka}{\textsc{Mokka}\xspace}
\newcommand{\marlin}{\textsc{Marlin}\xspace}
\newcommand{\geant}{\textsc{Geant4}\xspace}
\newcommand{\slic}{\textsc{SLIC}\xspace}
\newcommand{\lcsim}{\texttt{org.lcsim}\xspace}
\newcommand{\slicPandora}{\texttt{slicPandora}\xspace}
\newcommand{\pythia}{\textsc{Pythia}\xspace}
\newcommand{\whizard}{\textsc{Whizard}\xspace}
\newcommand{\pandora}{\textsc{PandoraPFA}\xspace}
\newcommand{\fastjet}{\textsc{FastJet}\xspace}
\newcommand{\tmva}{TMVA\xspace}
\newcommand{\roofit}{\texttt{RooFit}\xspace}

\title{Prospects for the Measurement of the Higgs Yukawa Couplings to b and c quarks, and muons at CLIC}
\author{Christian Grefe\thanksref{e1,addr1}
        \and
        Tom\'{a}\v{s} La\v{s}tovi\v{c}ka\thanksref{e2,addr2}
        \and
        Jan Strube\thanksref{e3,addr1} 
}

\thankstext{e1}{e-mail: christian.grefe@cern.ch}
\thankstext{e2}{e-mail: tomas.lastovicka@cern.ch}
\thankstext{e3}{e-mail: jan.strube@cern.ch}

\institute{CERN, Switzerland\label{addr1}
          \and
          Institute of Physics, Academy of Sciences of the Czech Republic, Prague, Czech Republic\label{addr2}
}

\date{Received: date / Accepted: date}
\maketitle
\begin{abstract}
The investigation of the properties of the Higgs boson, especially a test of the predicted linear dependence of the branching ratios on the mass of the final state is going to be an integral part of the physics program at colliders at the energy frontier for the foreseeable future. The large Higgs boson production cross section at a $\unit[3]{TeV}$ CLIC machine allows for a precision measurement of the Higgs branching ratios. The cross section times branching ratio of the decays \hbb, \hcc and \hmumu of a Standard Model Higgs boson with a mass of 120~GeV can be measured with a statistical uncertainty of 0.23\%, 3.1\% and 15\%, respectively, assuming an integrated luminosity of 2~\abinv.
\end{abstract}

\section{Introduction}
\label{sec:Introduction}
The Higgs mechanism of the Standard Model predicts the existence of a fundamental spin-0 particle. Recently, the ATLAS and CMS experiments at the LHC have observed a particle which is consistent with the predictions for a Standard Model Higgs boson, but its properties remain to be studied~\cite{ATLASHiggs, CMSHiggs}. In particular, the Standard Model predicts a linear dependence between the Higgs branching ratios to fermions and their mass. This relation could be altered by the presence of new physics. The detailed exploration of the Higgs sector is thus instrumental to our understanding of the fundamental interactions.
The compact linear collider (CLIC) is a proposed \epem collider with a maximum centre-of-momentum energy $\sqrt{s} = \unit[3]{TeV}$, based on a two-beam acceleration scheme~\cite{cdrvol1}. The Higgs boson production cross section of \unit[421]{fb} in the dominant \PW-fusion channel allows for precision measurements of the Yukawa couplings.
The beam of the \unit[3]{TeV} CLIC consists of bunch trains of 312 bunches, which are separated by \unit[0.5]{ns}. The small beam size and large electric field in the bunches, required to achieve the peak luminosity of $\unit[5.9\times 10^{34}]{cm^{-2}s^{-1}}$, lead to a large cross section of real and virtual two-photon processes that are a background to the processes of interest produced in the electron-positron collision. On average, 3.2 \gghadrons events are produced at every bunch crossing at \sqrts=\unit[3]{TeV}.

We present simulation studies of the measurements of the branching ratios \hbb, \hcc~\cite{lcd:2011-036} and \hmumu~\cite{lcd:grefeHmumu2011} at such a machine. These studies of the Higgs branching ratios are part of the benchmarking analyses presented in the CLIC Conceptual Design Report~\cite{cdrvol2}. They are carried out in a \geant{}-based simulation~\cite{Allison:2006ve} of the \clicsid~\cite{lcd:grefemuennich2011} detector concept, with full account of Standard Model backgrounds and using a realistic reconstruction in presence of \gghadrons background. The latter is reduced partly by removing hits that are out of time with the physics process, partly by advanced off-line reconstruction techniques.

\section{The \clicsid Detector Model}
\label{sec:DetectorModel}
The \clicsid detector model in which these studies are carried out is a general-purpose detector with a \unit[4]{$\pi$} coverage and is based on the SiD concept~\cite{Aihara:2009ad} developed for the ILC. It has been adapted~\cite{lcd:grefemuennich2011} to meet the specific detector requirements at CLIC. It is designed for particle flow calorimetry using highly granular calorimeters.

A superconducting solenoid with an inner radius of \unit[2.7]{m} provides a central magnetic field of \unit[5]{T}. The calorimeters are placed inside the coil and consist of a 30 layer tungsten--silicon electromagnetic calorimeter with \unit[$3.5\times3.5$]{mm$^2$} segmentation, followed by a tungsten--scin\-til\-lat\-or hadronic cal\-o\-rim\-e\-ter with 75 layers in the barrel region and a steel--scin\-til\-lat\-or hadronic calorimeter with 60 layers in the endcaps. The read-out cell size in the hadronic calorimeters is \unit[$30\times30$]{mm$^2$}. The iron return yoke outside of the coil is instrumented with nine double-RPC layers with \unit[$30\times30$]{mm$^2$} read-out cells for muon identification.

The silicon-only tracking system consists of five \unit[$20\times20$]{$\micron^2$} pixel layers followed by five strip layers with a pitch of \unit[25]{\micron}, a read-out pitch of \unit[50]{\micron} and a length of \unit[92]{mm} in the barrel region. The tracking system in the endcap consists of four stereo-strip disks with similar pitch and a stereo angle of 12\degrees, complemented by seven pixelated disks in the vertex and far-forward region at lower radii with pixel sizes of \unit[$20\times20$]{$\micron^2$}.

The forward region is instrumented with a LumiCal, with coverage down to \unit[40]{mrad}, and a BeamCal, with coverage down to \unit[10]{mrad}.

The trigger-less readout integrates over \unit[10]{ns} for all sub-detectors except the hadronic calorimeter, which has an integration time of \unit[100]{ns} to allow for shower development in the tungsten absorber. The silicon detectors allow time stamping of the recorded hits with a precision of a few ns.

\section{Analysis Framework}
\label{sec:Samples}
The physical processes are produced with the \whizard~\cite{Kilian:2007gr,Moretti:2001zz} event generator, taking into account the CLIC beam spectrum, with fragmentation and hadronisation handled by the \pythia~\cite{Sjostrand2006} package. The branching ratios of a \unit[120]{GeV} Standard Model Higgs boson are: $\mathrm{BR}(\hbb) = 6.48\times10^{-1}$, $\mathrm{BR}(\hcc) = 3.27\times10^{-2}$ and $\mathrm{BR}(\hmumu) = 2.44\times10^{-4}$~\cite{Denner:2011mq}. The events are simulated in the \clicsid detector model using \slic~\cite{Graf:2006ei}, which is a thin wrapper around \geant. They are reconstructed by the \lcsim and \slicPandora packages. Unlike in analyses at lower-energy linear colliders, which use DURHAM-style jet finders that operate on all particles in the event, it was found that the beam-jets of algorithms originally developed for hadron colliders, lead to a crucial improvement of the jet-energy resolution and reduce the effect of the forward-peaking \gghadrons events greatly. In the analysis of Higgs decays to b and c quarks, we use the $k_t$ algorithm~\cite{Ellis:1993tq} as implemented by the \fastjet~\cite{Cacciari:2011ma,Cacciari:2005hq} package. The LCFI~\cite{lcfi} package is used for flavour tagging. The assumed luminosity of the analyses is \unit[2]{\abinv}, corresponding to about 4 years of data taking at nominal conditions, assuming 200 days of running per year at an efficiency of 50\%.
\begin{table}
 \centering
\begin{tabular}{r r r}
\toprule
Process & $\sigma$ (fb) & Short label \\
\midrule
$\epem \to \PH \nuenuebar$; \hmumu    & 0.120  & \hmumu    \\
$\epem \to \PH \nuenuebar$; \hbb &   272   & \hbb  \\
$\epem \to \PH \nuenuebar$; \hcc &   13.7  & \hcc \\
\midrule
$\epem \to \mpmm \nunubar$                       & $132^{*}$          &  $\mpmm \nunubar$    \\
$\epem \to \mpmm \epem$                          & $346^{*}$          &  $\mpmm \epem$       \\
$\epem \to \mpmm$                                &  $12^{*}$ 	   &    $\mpmm$             \\
$\epem \to \tptm$                                & $250^{*}$          & $\tptm$             \\
$\epem \to \tptm \nunubar$                       & $125^{*}$          & $\tptm \nunubar$    \\
$\epem \to \qq$								     & 3100	&    $\qq$				  \\
$\epem \to \qq\nunubar$						     & 1300	 & $\qq\nu\nu$	      \\
$\epem \to \qq\epem$							     & 3300	&  $\qq\epem$			  \\
$\epem \to \qq \Pe \nu$						     & 5300	&  $\qq \Pe \nu$          \\
{\scriptsize generator level:}\hfill $\gamgam \to \mpmm$        & $20000^{*}$       & $\gamgam \to \mpmm$ \\
\bottomrule
\end{tabular}
 \caption{List of processes considered for this analysis with their respective cross section $\sigma$ and the number of simulated events $N_{\mathrm{events}}$. The cross section takes into account the CLIC luminosity spectrum. Cross sections marked with * include a cut on the invariant mass of the muon pair to lie between 100 and 140 GeV.}
 \label{tab:samples}
\end{table}

\subsection{Rejection of \gghadrons backgrounds}
A \unit[3]{TeV} CLIC produces 3.2 \gghadrons events per bunch crossing on average. The spacing of \unit[0.5]{ns} between bunches leads to pile-up in the subdetectors, which integrate over multiple bunch crossings. Identifying the time of the physics event and reading out only a window of \unit[10]{ns} for the subdetectors, except for the barrel of the hadronic calorimeter, for which \unit[100]{ns} are read out, reduces the the number of \gghadrons events in the data sample by about a factor of 15.

To take into account the effect of this background on the measurement, a sample of events from \gghadrons corresponding to 60 bunch crossings is mixed with each physics event for the analysis of the Higgs decaying to \PQb and \PQc quarks. In the \hmumu analysis, only the signal sample was mixed with events from \gghadrons background. These events are also simulated in the \geant model of the \clicsid detector. The equivalent of 60 bunch crossings is a compromise between realistic description and computational constraints. The \gghadrons events are forward-peaking; they are described in more detail elsewhere~\cite{lcd:2011-DannheimSailerBgrNote}. Their contribution to hits in the barrel hadronic calorimeter, which, in principle, accumulates the equivalent of up to 200 bunch crossings, is small.
Table~\ref{tab:samples} lists the physics processes that were taken into account in the analyses.

In addition to applying read-out windows off-line, the computation of the cluster time allows to further reduce this background. Assuming ns precision of the calorimeter hit times results in sub-ns precision for the cluster time, which is calculated as a truncated mean of the corresponding hit times. The production time of the reconstructed particle is obtained by correcting the cluster time for its time of flight through the magnetic field. The production time of the particle is required to be consistent with the start of the physics event. Consistency is defined by a time window, whose size depends on the type of particle (hadronic or electromagnetic), its momentum and polar angle $\theta$. This reduces the energy from \gghadrons processes to the event further by a factor of 6 or more, while only about 0.5\% of the energy from a typical physics event is removed~\cite{LCD:2011-28}.

\section{Measurement of Higgs decays to pairs of b and c quarks}
\begin{figure}
    \subfloat[]{\includegraphics[width=.5\columnwidth]{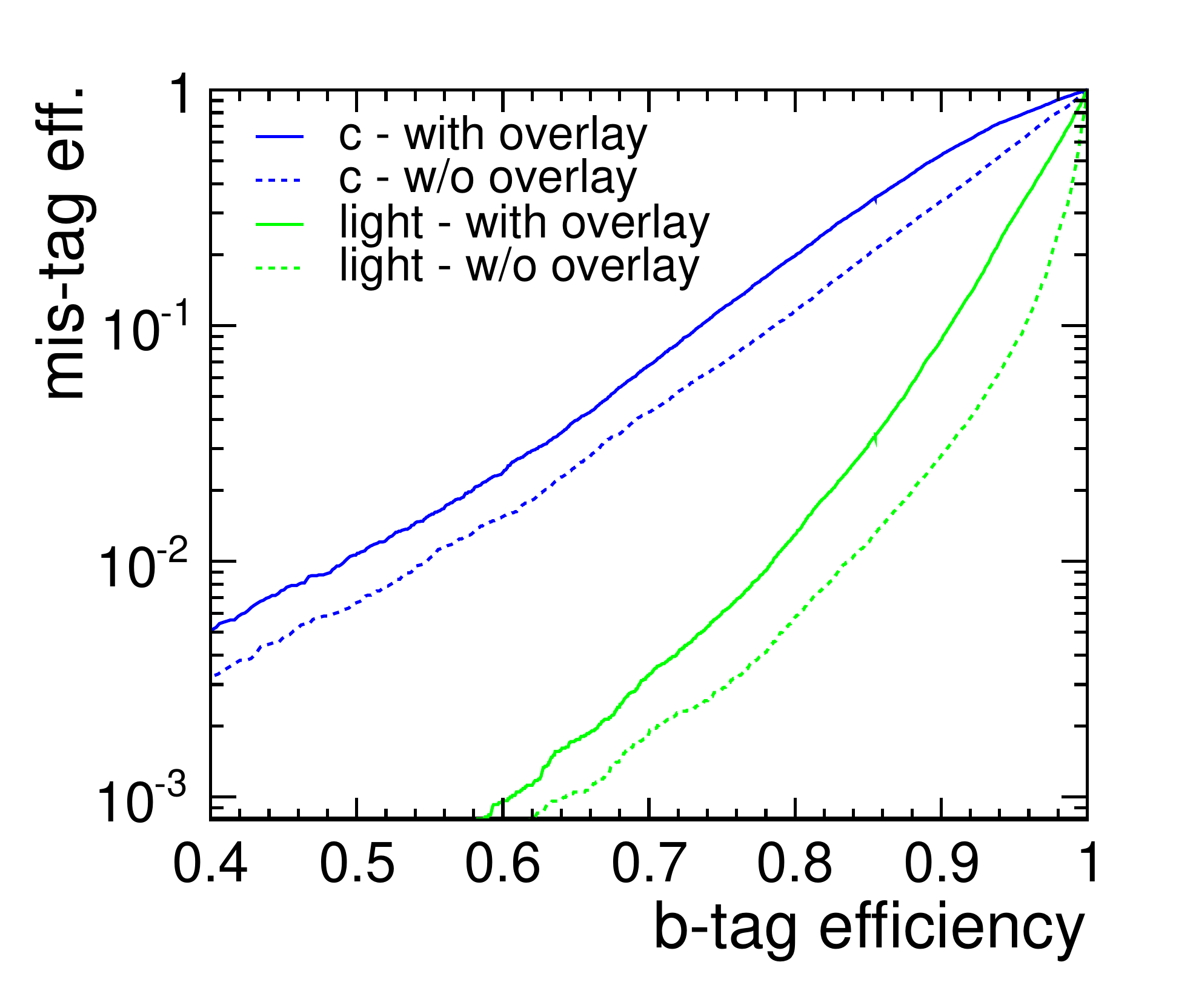}} \hfill
    \subfloat[]{\includegraphics[width=.5\columnwidth]{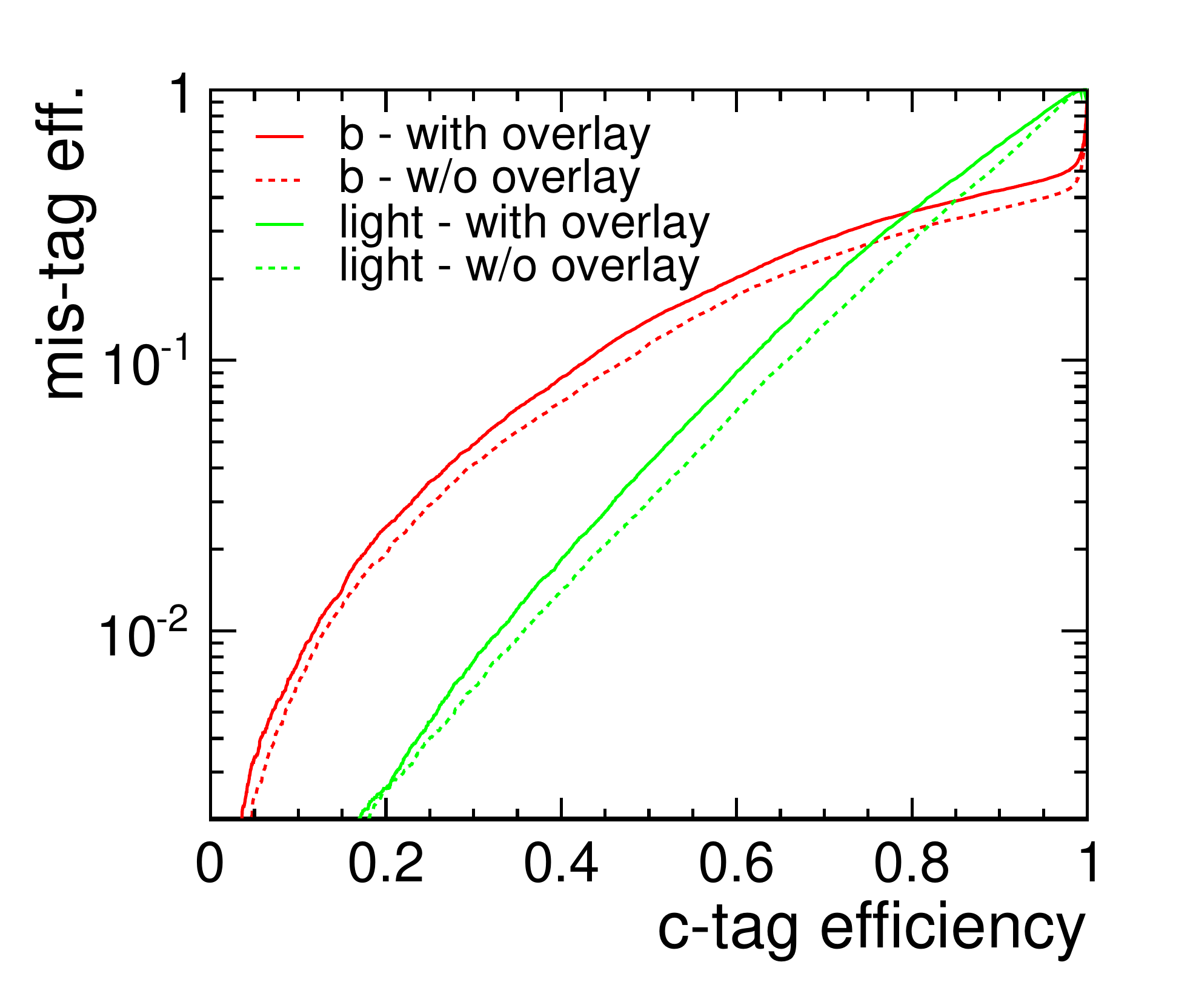}}
    \caption{Left: Efficiency of tagging a \PQb-jet as \PQc (blue lines) or light (green lines) versus tagging it as \PQb. Right: Efficiency of tagging a \PQc-jet as \PQb (red lines) or light (green lines) versus tagging it as \PQc. The solid lines show performance with pile-up of \gghadrons events, the dashed lines without this background.}
    \label{fig:flavorTaggingPerformance}
\end{figure}

\begin{figure}
	\subfloat[\hbb]{\includegraphics[width=.5\columnwidth]{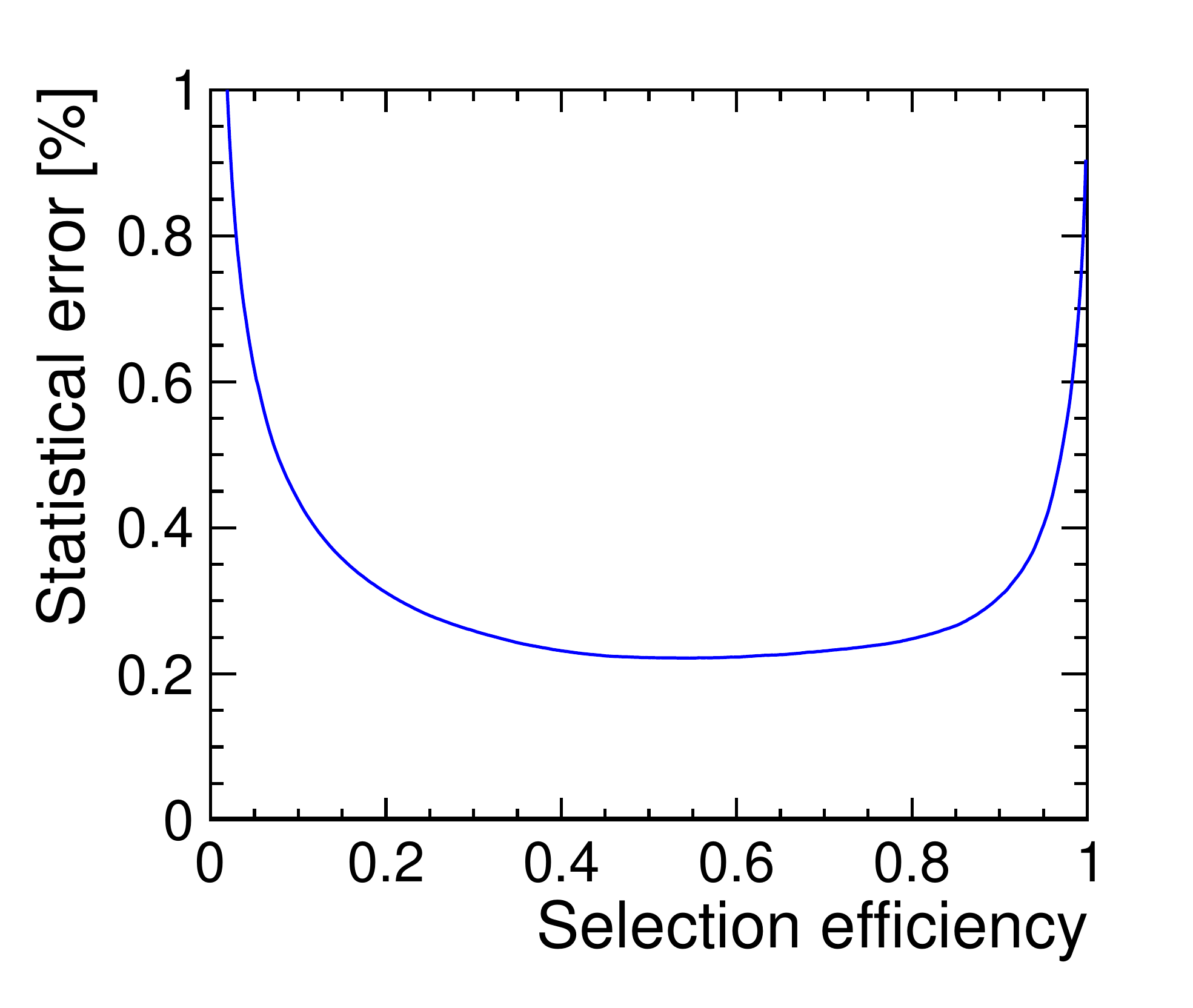}} \hfill
	\subfloat[\hcc]{\includegraphics[width=.5\columnwidth]{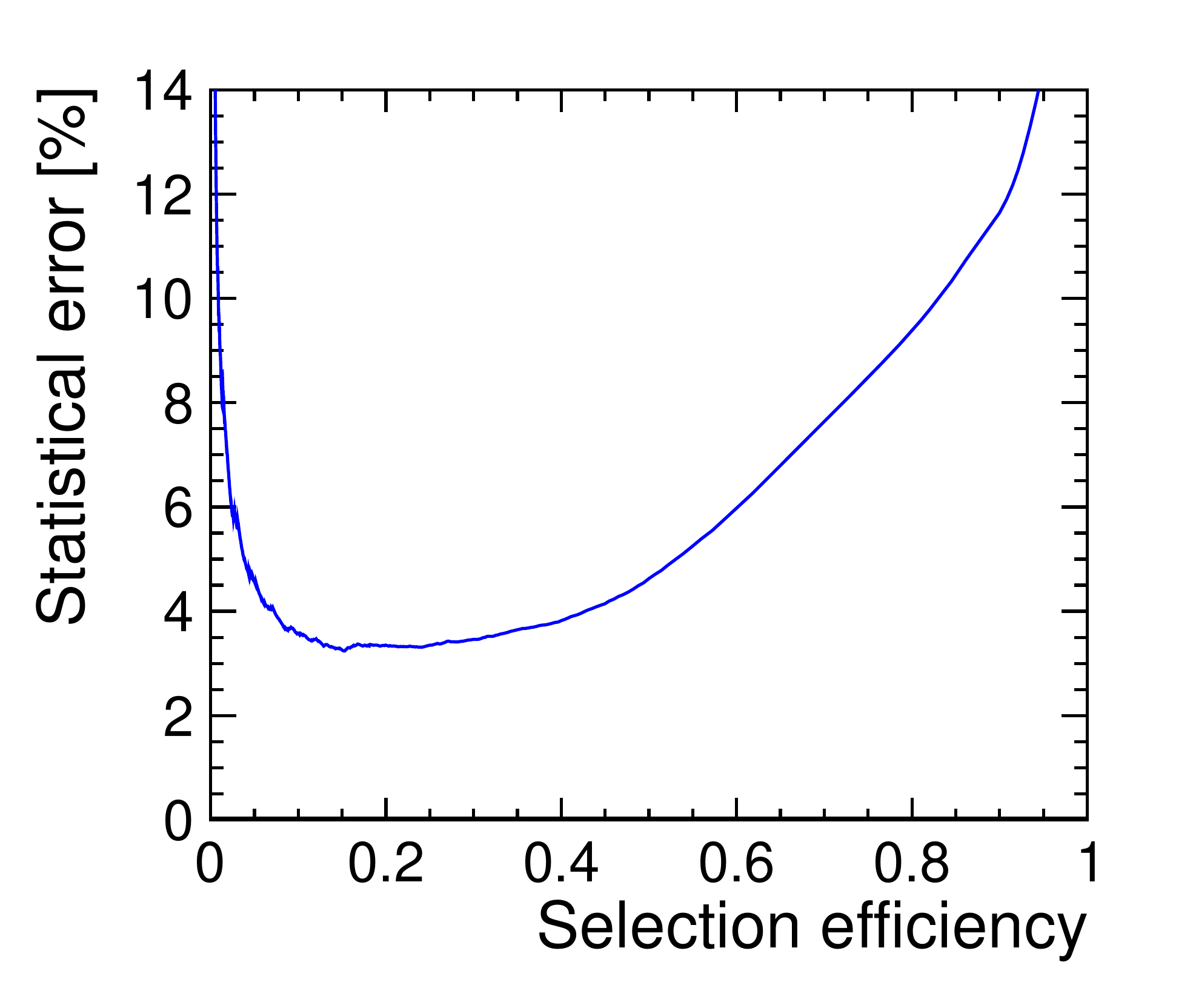}}
	\caption{Statistical uncertainty of the measurement of cross section times branching ratio versus selection efficiency of the neural network. Left: The neural network was trained to identify \hbb decays from di-jet backgrounds including \hcc. Right: The neural network was trained on \hcc as signal and di-jets including \hbb as background.}
	\label{fig:bcSelectionEfficiency}
\end{figure}

The particles passing the pre-selection based on the reconstructed production time are clustered into two jets using the $k_t$ algorithm as implemented in the \fastjet package.
The LCFI flavour tagging package finds secondary vertices in each jet and uses them, along with complementary track-based information, in a neural network to distinguish \PQb-, \PQc-, and light quark jets.
Figure~\ref{fig:flavorTaggingPerformance}(a) shows the mis-tag rate for \PQc-jets and light jets as \PQb-jets versus the \PQb-tag efficiency, while Figure~\ref{fig:flavorTaggingPerformance}(b) shows the mis-tag rate for \PQb-jets and light jets as \PQc-jets versus the \PQc-tag efficiency. The presence of \gghadrons background is found to reduce the flavour tagging performance. This effect is correlated with degraded jet finding quality due to the \gghadrons background. For instance, at the \PQb-tag efficiency of 70\% the mis-tag rate for \PQc-jets (light jets) increases from 4.3\% (0.19\%) without overlay to 6.8\% (0.33\%) with overlay.
The main SM background of the measurement of the decays \hbb and \hcc is from two-jet processes $\epem \to \qq\nunubar$, due to their large cross section, and from processes with two measured jets and additional particles that escape detection.
The invariant mass of the jet pair is the major discriminant between decays of Higgs and of \PZ bosons. It is used in a second neural network, together with the output of the b-flavour-tagging network and the following variables:
\begin{itemize}
\item The maximum of the absolute values of jet pseudorapidities.
\item The sum of the remaining LCFI jet flavour tag values, i.e. \PQc-tag against \PQu\PQd\PQs\PQb-background, \PQc-tag against \PQb-background and \PQb-tag against uds-background.
\item $R_{\eta\phi}$, the distance of jets in the $\eta-\phi$ plane.
\item The sum of jet energies.
\item The total number of leptons in an event.
\item The total number of photons in an event.
\item The acoplanarity of the jets.
\end{itemize}
The neural network selection efficiency $S/S_{total}$ versus the statistical uncertainty $\sqrt{S+B}/S$ on the measurement of the number of signal events $S$ and background events $B$ is shown in Figure~\ref{fig:bcSelectionEfficiency} for the two neural networks that were trained on \hbb and \hcc as signal, respectively. The optimal selection is at the local minimum of the curve, at a selection efficiency of 55\% for \hbb with a sample purity of 65\%, corresponding to a statistical uncertainty of 0.23\%. The optimal selection for \hcc has an efficiency of 15\%, corresponding to a sample purity of 24\% and a statistical uncertainty of 3.1\%. Purity values reflect the fact that \PQb-jets can be distinguished from \PQc-jets with high purity, while incompletely reconstructed \PQb-jets make up a large fraction of the background to \PQc-jet selection, making the analysis more challenging.

\section{Measurement of Higgs decays to pairs of muons}
 The measurement of the rare decay \hmumu requires high luminosity operation and sets stringent limits on the momentum resolution of the tracking detectors. The branching ratio of the decay of a Standard Model Higgs boson to a pair of muons is important as the lower end of the accessible decays and defines the endpoint of the test of the predicted linear dependence of the branching ratios to the mass of the final state particles.

\subsection{Event Selection}
\label{sec:EventSelection}
The average muon reconstruction efficiency for polar angles greater than 10\degrees is 99.6\% without \gghadrons background. When adding this background the muon reconstruction efficiency deteriorates to 98.4\% in this region of polar angles. The efficiency for smaller polar angles is limited by the acceptance of the tracking detectors. The events are required to have at least two reconstructed muons, each with a transverse momentum of more than \unit[5]{GeV}. In case there are more than two muons reconstructed, the two most energetic ones are used, which are referred to as $\upmu_1$ and $\upmu_2$. In addition, the invariant mass of the two muons $M(\mumu)$ is required to be between \unit[105]{GeV} and \unit[135]{GeV}. The total reconstruction efficiency of the signal sample is 72\% in the presence of \gghadrons background. The inefficiency is dominated by acceptance effects.

\begin{figure}
  \centering
  \includegraphics[width=.75\linewidth]{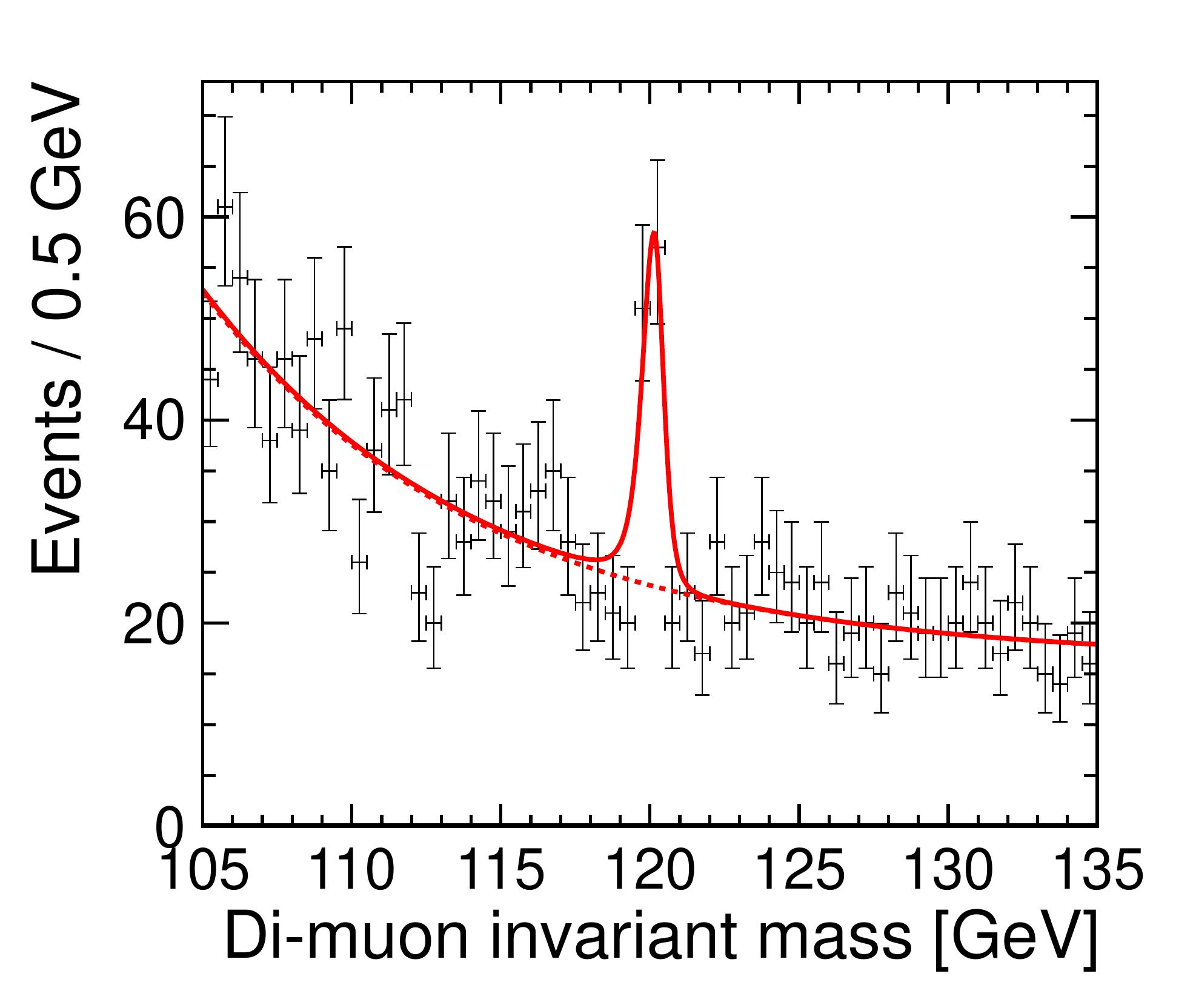}
  \caption{Maximum Likelihood fit of the Higgs mass in the data sample after selection cuts}\label{fig:mumu_massFit}
\end{figure}

The event selection is done using the boosted decision tree (BDT) classifier implemented in \tmva~\cite{TMVA:2010}. The $\mpmm$, $\tptm$ and $\tptm \nunubar$ samples are not used in the training of the BDT, but are effectively removed by the classifier nevertheless. The variables used for the event selection by the BDT are:
\begin{itemize}
 \item The visible energy excluding the two reconstructed muons $E_{\mathrm{vis}}$.
 \item The scalar sum of the transverse momenta of the two muons $\pT(\upmu_1) + \pT(\upmu_2)$.
 \item The helicity angle $\cos\theta^*(\mumu) = \frac{\vec{p}'(\upmu_1) \cdot \vec{p}(\mumu)}{|\vec{p}'(\upmu_1)| \cdot |\vec{p}(\mumu)|}$, where $\vec{p}'$ is the momentum in the rest frame of the di-muon system.
 \item The relativistic velocity of the di-muon system $\upbeta(\mumu)$, where $\upbeta = \frac{v}{c}$.
 \item The transverse momentum of the di-muon system $\pT(\mumu)$.
 \item The polar angle of the di-muon system $\theta(\mumu)$.
\end{itemize}
The most powerful variable to distinguish signal from background events is the visible energy whenever there is an electron within the detector acceptance. Otherwise the background can be reduced by the transverse momentum of the di-muon system or the sum of the two individual transverse momenta.
Figure~\ref{fig:mumu_massFit} clearly shows the Higgs peak in the invariant mass distribution after the event selection.

\subsection{Invariant Mass Fit}
\label{sec:MassFit}	
The number of signal events is determined by an unbinned maximum likelihood fit of the invariant mass distribution of the combined signal and background sample. This sample is randomly selected from all simulated events, according to the assumed integrated luminosity of \unit[2]{\abinv}. The expected shapes of the signal and background contributions are determined from a fit to the full statistics of the respective sample. The distribution of the invariant mass in the $\epem \to \hmumu$ sample has a tail towards lower masses because of final state radiation. It is described by two half Gaussian distributions, each with an exponential tail.
The background is well described by an exponential parametrisation, obtained from a background-only sample.

The BDT selection with the highest signal significance yields a total signal selection efficiency of 21.7\%, corresponding to about 53 selected events in \unit[2]{\abinv}. The relative statistical uncertainty on the cross-section times branching ratio obtained from the fit of the invariant mass distribution is 26.3\%. This corresponds to a signal significance of approximately $3.8\sigma$. Without addition of the \gghadrons background the relative statistical uncertainty on the cross-section times branching ratio improves to 23\%, due to higher signal selection efficiency.

\subsection{Study of the Momentum Resolution}
\label{sec:MomentumResolution}
\begin{table}
 \centering
\begin{tabular}{r r r}
\toprule
$\sigma(\Delta\pT)/\pT^2$    & $\sigma(\Delta M(\mumu))$ & Stat. \\
& & uncertainty \\
\midrule
\unit[$10^{-3}$]{GeV$^{-1}$} & \unit[6.5]{GeV}           &   -               \\
\unit[$10^{-4}$]{GeV$^{-1}$} & \unit[0.70]{GeV}          & 34.3\%            \\
\unit[$10^{-5}$]{GeV$^{-1}$} & \unit[0.068]{GeV}         & 18.2\%            \\
\unit[$10^{-6}$]{GeV$^{-1}$} & \unit[0.022]{GeV}         & 16.0\%            \\
\bottomrule
\end{tabular}
 \caption{Dependence of the statistical uncertainty of the measurement of cross section times branching ratio for the decay $\PSh \to \mpmm$ on the momentum resolution $\sigma(\Delta\pT)/\pT^2$. The study assumes an integrated luminosity of 2~\abinv. The values do not include the impact of the \gghadrons background and the possible reduction of the $\epem \to \mpmm \epem$ background using electron tagging in the forward calorimeters.}
 \label{tab:resultsMomentumResolution}
\end{table}
The ability to measure the decay \hmumu depends crucially on the momentum resolution of the tracking detectors. In a fast simulation study, different values for the momentum resolution were applied to the true muon momenta. For each assumed momentum resolution an individual BDT was trained to optimise the event selection for the invariant mass fit, which is performed as described above. For this study the impact of the \gghadrons background was neglected. The results are shown in Table~\ref{tab:resultsMomentumResolution}. We find an average resolution of at least $5\times\unit[10^{-5}]{GeV^{-1}}$ is required in order for the momentum resolution not to be the dominant uncertainty contribution in a \unit[2]{\abinv} measurement of the decay \hmumu. The average momentum resolution in the fully simulated \hmumu sample is $4\times\unit[10^{-5}]{GeV^{-1}}$. The results from the fast simulation study are thus consistent with those found Section~\ref{sec:MassFit}.

\subsection{Forward Electron Tagging}
\begin{figure}
  \centering
  \includegraphics[width=.49\linewidth]{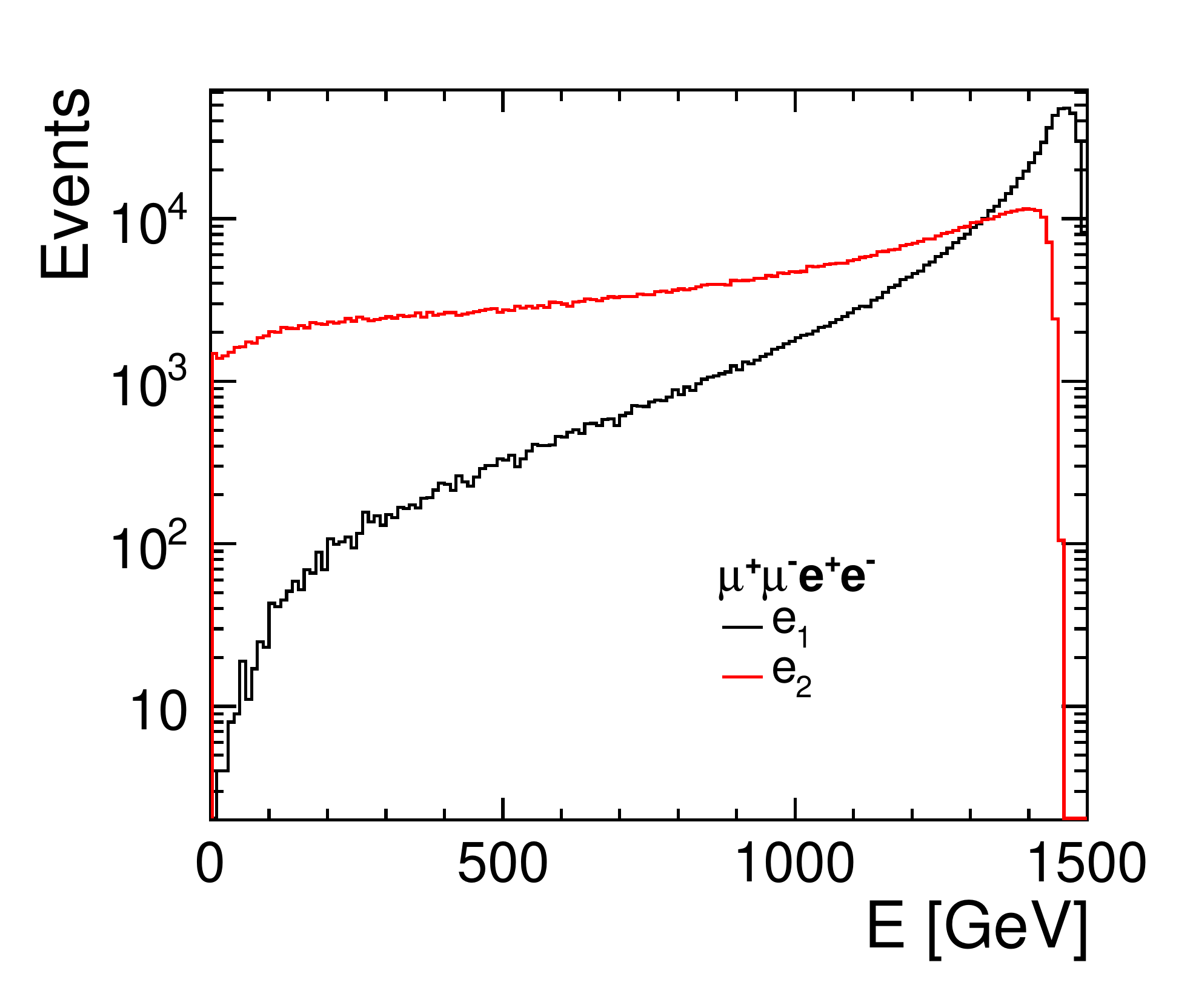} \hfill
  \includegraphics[width=.49\linewidth]{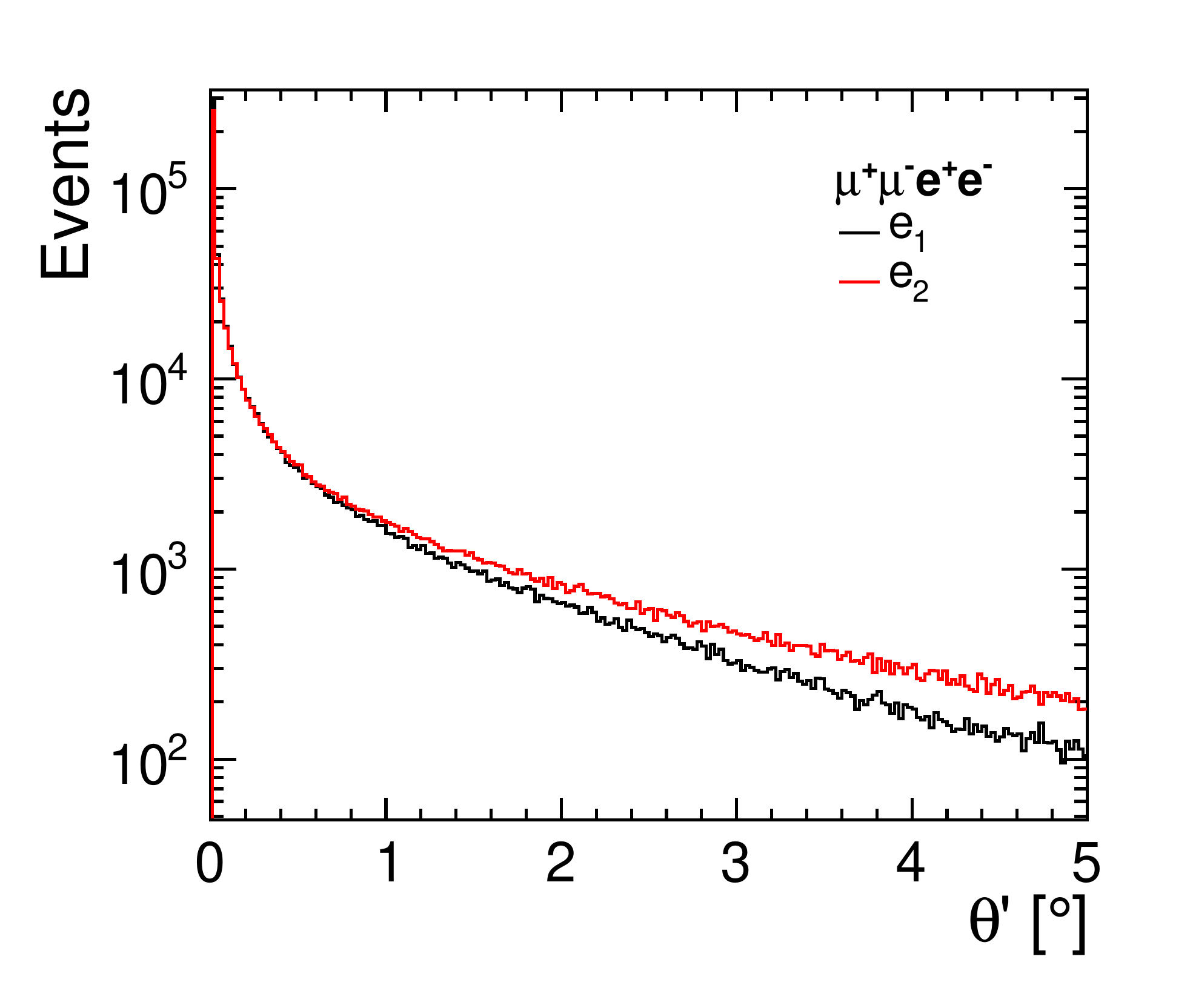}
  \caption{Kinematic distributions of the most and the second most energetic electron in $\mpmm\epem$ events. Left: Distribution of the simulated electron energy. Right: Distribution of the polar angle $\theta^{'}$ with respect to the outgoing beam axis.}
  \label{fig:mumuee_kinematicDistributions}
\end{figure}

The dominant contributions to the reducible background are from \PZ pair production, where one Z decays to a pair of muons and the other decays invisibly, and from the t-channel diagram contributing to $\epem \to \mpmm \epem$. In the latter the electron-positron pair goes in the very forward direction. We have investigated a possible reduction of this background using the forward calorimeters LumiCal and BeamCal. The distributions of energy and angle with the outgoing beam axis of the most and second most energetic electrons in $\epem \to \mpmm \epem$ events are shown in Figure~\ref{fig:mumuee_kinematicDistributions}. Although most electrons are produced at very low polar angle, a large fraction of the electrons are within the fiducial volumes of the LumiCal and BeamCal, which have an acceptance of 44~mrad and 15~mrad, respectively, with respect to the outgoing beam axis. Since the forward calorimeters were not part of the full detector simulation, we assume ad-hoc electron tagging efficiencies in these two calorimeters to reject background events. Afterwards, a dedicated BDT classifier is trained on the pre-selected background samples using the variables described in Section~\ref{sec:EventSelection}. For example, assuming an electron tagging efficiency of 95\% in the LumiCal improves the total signal selection efficiency to 49.7\%, which results in a statistical uncertainty on the cross-section times branching ratio measurement of 15.7\%. Assuming a higher electron tagging efficiency of 99\% in the LumiCal improves this result to approximately 15\%. If the BeamCal is used in addition, assuming an average electron tagging efficiency of 70\% in its fiducial volume, the statistical uncertainty can be improved to 14.5\%.

An independent study~\cite{sailerphd} of the electron tagging efficiency in the forward calorimeters at a CLIC detector, taking into account the \gghadrons background as well as \epem-pair background, confirms the efficiencies we assume here.

\section{Results}
\label{sec:result}
We have demonstrated the potential of measuring the cross section times branching ratios of a 120 GeV Higgs boson at a 3 TeV CLIC with high precision.
For the measurement of Higgs decays to quarks, 0.23\% and 3.1\% statistical uncertainty can be achieved for the decays \hbb and \hcc, respectively. This includes the effect of background from \gghadrons on the flavour tagging. Given the experience of the LEP experiments~\cite{ALEPH:2005ab} in the measurements of hadronic Z decays, with systematic uncertainties between 0.3\% - 1.2\% for $R^{0}_{\PQb}$ and between 1.2\% and 10\% for $R^{0}_{\PQc}$, one can assume that a systematic uncertainty of around 1\% is achievable in \hbb and around 5\% in \hcc. 

For the rare decay \hmumu, the cross section times branching ratio can be measured to a precision of about 15\% if the background from $\epem \to \mpmm\epem$ can be reduced using tagging of electrons in the LumiCal with an efficiency of 95\%, and the average momentum resolution is not worse than $5\times 10^{-5}$. The effect of background from \gghadrons has been taken into account. From the measurements of the branching ratio of Z decays to a pair of muons at the LEP experiments, with systematic uncertainties between 0.1 and 0.4\%, depending on the experiment, one can assume that the systematic uncertainties related to detector effects are of the order of 1\% or less. The expected uncertainty of the peak luminosity is currently being studied but is estimated to be around 1\% or less.

\subsection{Extracting the Higgs Coupling Constants}
The uncertainties on the measurements of cross section times branching fraction can be translated to an uncertainty on the coupling constants. A global fit~\cite{Desch:2001xh} to the complete set of measured electroweak observables gives the most accurate picture of the nature of the coupling constants. In absence of the full set of measurements, we estimate the achievable precision on the Higgs couplings in the measured channels by assuming that deviations from the Standard Model parameters occur only in the channel under consideration~\cite{Wells15Nov2011}. Using a recent overview of the uncertainties of the Standard Model Higgs branching ratios~\cite{Denner:2011mq}, Table~\ref{tab:couplingConstants} summarises conservative estimates on the achievable sensitivity to Standard Model Higgs coupling constants. For the hadronic decays, even the combination of the statistical uncertainty and a conservative average of the systematic uncertainties from similar measurements at LEP, as discussed above, is dominated by the current theoretical uncertainties of 2.8\% for \hbb and 12.2\% for \hcc.
In the case of \hmumu the statistical uncertainties will dominate both the systematic uncertainties and the current theoretical uncertainty of 6.4\%. 
\begin{table}[h]
  \caption{Statistical uncertainties of the measurements of the cross section times branching fraction, and the sensitivity to the SM b, c and $\mu$ Yukawa Higgs coupling constants at a \unit[3]{TeV} CLIC with an integrated luminosity of \unit[2]{\abinv}}
  \label{tab:couplingConstants}
  \centering
  \begin{tabular}{lcc}
  \toprule
      & $\sigma B$ statistical  & Sensitivity to \\
      & uncertainty (\%) & SM Yukawa deviation (\%)\\
  \midrule
    \hbb & 0.23 & 4 \\
    \hcc & 3.1 & 6 \\
    \hmumu & 15 & 7.5 \\
  \bottomrule
  \end{tabular}
\end{table}
\section{Acknowledgements}
The authors would like to thank Stephane Poss for generating the event samples and managing the simulation and reconstruction on the grid.
\bibliographystyle{unsrt}
\bibliography{Higgs_BR}




\end{document}